\documentclass{aa}
\usepackage{graphicx}
\usepackage{multirow} 
\usepackage{natbib}
\bibpunct{(}{)}{;}{a}{}{,}

\normalfont

\newcommand{\Msun}{~M_\odot}
\newcommand{\lsim}{\raise0.3ex\hbox{$<$}\kern-0.75em{\lower0.65ex\hbox{$\sim$}}}
\newcommand{\gsim}{\raise0.3ex\hbox{$>$}\kern-0.75em{\lower0.65ex\hbox{$\sim$}}}

\newcommand{\eg}{{e.g.\/\ }}
\newcommand{\ie}{{i.e.,\/\ }}

 \defcitealias{CF94}{CF94}
 \defcitealias{FLC96}{FLC96}
 \defcitealias{Nymark06}{NFK06}

\begin{document}         

\title{Modeling the X-ray emission of SN~1993J}
\author{Tanja K. Nymark\inst{1}
\and Poonam Chandra\inst{2,3}
\and Claes Fransson\inst{1}}
\titlerunning{}
\authorrunning{Nymark et al.}
\institute{Department of Astronomy, Stockholm University, AlbaNova University Center, SE-106 91 Stockholm, Sweden
\and
Jansky Fellow, National Radio Astronomy Observatory
\and
Department of Astronomy, University of Virginia, P.O. Box 400325, Charlottesville, VA 22904-4325}
\offprints{Tanja K. Nymark, \email{tanja@astro.su.se}}

\date{Received  / Accepted }

\abstract
{}
{We investigate the effects of radiative shocks on the observed X-ray emission from the Type~II supernova SN1993J. To this end, the X-ray emission is modeled as a result of the interaction between the supernova ejecta and a dense circumstellar medium at an age of 8 years.}
{The circumstances under which the reverse shock is radiative are discussed and the observed X-ray emission is analyzed using the numerical code described in Nymark et al (2006).
We argue that the original analysis of the X-ray observations suffered from the lack of self-consistent models for cooling shocks with high density and velocity, leading to questionable conclusions about the temperatures and elemental abundances. We reanalyze the spectra with our numerical model, and discuss the expected spectra for different explosion models for the progenitors. }
{We find that the spectra of SN~1993J are compatible with a CNO-enriched composition and that  the X-ray flux is dominated by the reverse shock.}
{}


\maketitle

\section{Introduction}
\label{sect:intro}
Many core-collapse supernovae (SNe) show strong X-ray and radio emission. This is, in particular, true for the Type~IIL, IIn and IIb SNe \citep[see \eg ][ for a review]{Immler03}.  We now know that this is a result of the interaction of the supernova ejecta and the circumstellar medium (CSM) of the supernova. A two-shock structure is formed consisting of the outgoing supernova shock (hereafter the forward shock) and a reverse shock, which moves backward into the ejecta \citep[][ hereafter~\citetalias{CF94}]{Chevalier1982a,Chevalier1982b, CF94}. The region behind the forward shock consists of shocked circumstellar material, with $T\sim 10^8-10^9$~K, while the region behind the reverse shock, which consists of shocked ejecta, is cooler, with $T\sim 10^7$~K.  The radio emission is caused by synchrotron emission from the forward shock \citep[\eg ][]{Sramek03}, while the X-ray emission originates partly from free-free and inverse Compton emission behind the forward shock, and partly from line emission in the cooler gas behind the reverse shock. Both the radio and the X-ray emission are powerful tools for understanding the physics of both the SNe and the CSM.

With the new generation of X-ray telescopes, XMM-Newton and Chandra, more detailed spectra have been obtained than ever before. This puts greater demands on the tools used for analyzing the spectra.   Traditionally, X-ray spectra have been analyzed with the XSPEC package, which includes a number of models that can be combined to get the best fit. In \citet[][ hereafter \citetalias{Nymark06}]{Nymark06}  we showed, however, that the SNe having the strongest X-ray emission are likely to have radiative shocks. This has not been accounted for in the analysis of the SNe observed in X-rays. Furthermore, none of the models included in the XSPEC distribution offers the possibility of fitting cooling shocks in a consistent manner. 

In radiative shocks a wide range of temperatures contribute to the spectrum. The cooling also affects the hydrodynamic structure and influences the emitting volume, and thus the total luminosity from the interaction region. This is not properly accounted for  when fitting with one, or even several, single-temperature models, and the elemental abundances can be over- or under-estimated in the resulting analysis. In \citetalias{Nymark06} we described a numerical model that computes the emission from a cooling shock self-consistently. This was applied to the coronal emission from the ring collision in SN~1987A by \citet{Groningsson06}. Here we apply this model to the X-ray spectra of SN1993J, obtained with Chandra \citep{Swartz03} and XMM-Newton \citep{ZimAsch03}.

\section{Observations and data analysis}
SN 1993J was discovered on March 28, 1993 in M82 \citep{Rip93}. It was one of the brightest supernovae of the twentieth century, and has been followed in great detail in many wavelengths in the decade following the explosion. On the basis of the detection of hydrogen lines in the spectrum SN1993J was first identified as a type~II supernova, but as the Balmer lines soon weakened its spectrum came to resemble that of a type Ib supernova, and it was re-classified as a Type~IIb
 \citep{FilMathHo93}. This transition is explained by an almost complete loss of the hydrogen envelope due to heavy pre-supernova mass loss, probably due to a binary interaction \citep[\eg][]{Nomoto93, Pods93, Ray93, Woosley94}. This hypothesis was confirmed by \citet{Maund04},
when they detected the unambiguous signature of a massive star, \ie the binary companion of the supernova progenitor, in HST observations of the supernova 10 years after the explosion.  
Evidence of circumstellar interaction was seen at early times in radio \citep{vanDyk94}, UV \citep{Fransson94}, X-rays \citep{Zim94, Uno02} and $\gamma $-rays by OSSE on CGRO \citep{Leising94}. The detection of the supernova in the radio \citep{Chandra04, Weiler07} and X-rays bands \citep{Uno02} even after a decade shows the presence of an extended dense CSM. 

During the first weeks ROSAT and ASCA observations showed a hard X-ray spectrum. This was confirmed by OSSE, which showed a hard spectrum, well fit by a free-free spectrum with $kT\sim 100$~keV \citep{Leising94}. However, after six months ASCA observations showed  that this  had changed to a soft spectrum \citep{Uno02}. This transition is consistent with circumstellar interaction, as the decreasing optical depth in the expanding cool shell allows more of the soft emission to escape (\citet{FLC96}, hereafter \citetalias{FLC96}).

In May 2000 SN~1993J was observed with Chandra \citep{Swartz03}. Strong Fe~L emission is evident in the spectrum, as are emission lines from \ion{Mg}{xi-xii} and \ion{Si}{xiii-xiv}.  \citet{Swartz03} fit the spectrum with three components -- two low temperature components ($kT\sim 0.35$ keV and $kT\sim 1.01$ keV) with no He, C, O and Ne but enhanced in N, Mg, Si and Fe; as well as a high temperature component ($kT\sim 6.0$ keV), consistent with solar abundances 
\citep{Swartz03}. This was argued to be consistent with an interaction model, in which the high temperature component comes from the circumstellar shock, while the two low-temperature components come from different parts of a radiative reverse shock. 

In April 2001 \citet{ZimAsch03} observed SN~1993J with XMM-Newton. Their spectrum was found to be best fit with two components (with $kT\sim 0.34$ keV and $kT\sim 6.4$ keV). These are in good agreement with the high and low temperature components of the Chandra spectrum, but the intermediate component found by \citet{Swartz03} in the Chandra spectrum is missing in the analysis of the XMM spectrum by \citet{ZimAsch03}. This could be caused by a real change in the spectrum between the two observations. Also here the two components are taken as evidence of a two-shock structure, but they found the abundances to be poorly constrained, and no conclusions were drawn about the composition of the ejecta. 

As we argue below, both these analyses are doubtful if the shock is radiative, as it most likely is. The abundances were kept as completely free parameters, and it is not clear how unique the solution is, especially with a realistic temperature model. The thin hydrogen envelope means that the reverse shock will traverse the hydrogen rich region on a short timescale, and the emission will already after a year be dominated by processed material, which gives us an excellent opportunity for testing stellar evolution theories. In this paper we take a different approach and use abundance models based on self-consistent stellar evolution models as input.

For the analysis in this paper we make new extractions of both the Chandra and  XMM-Newton data to fit with our
models. Here we describe the observational and data analysis procedure.

\subsection{The Chandra dataset}

\begin{figure*}[ht]
\begin{center}
\resizebox{0.4\hsize}{!}{\rotatebox{0}{\includegraphics{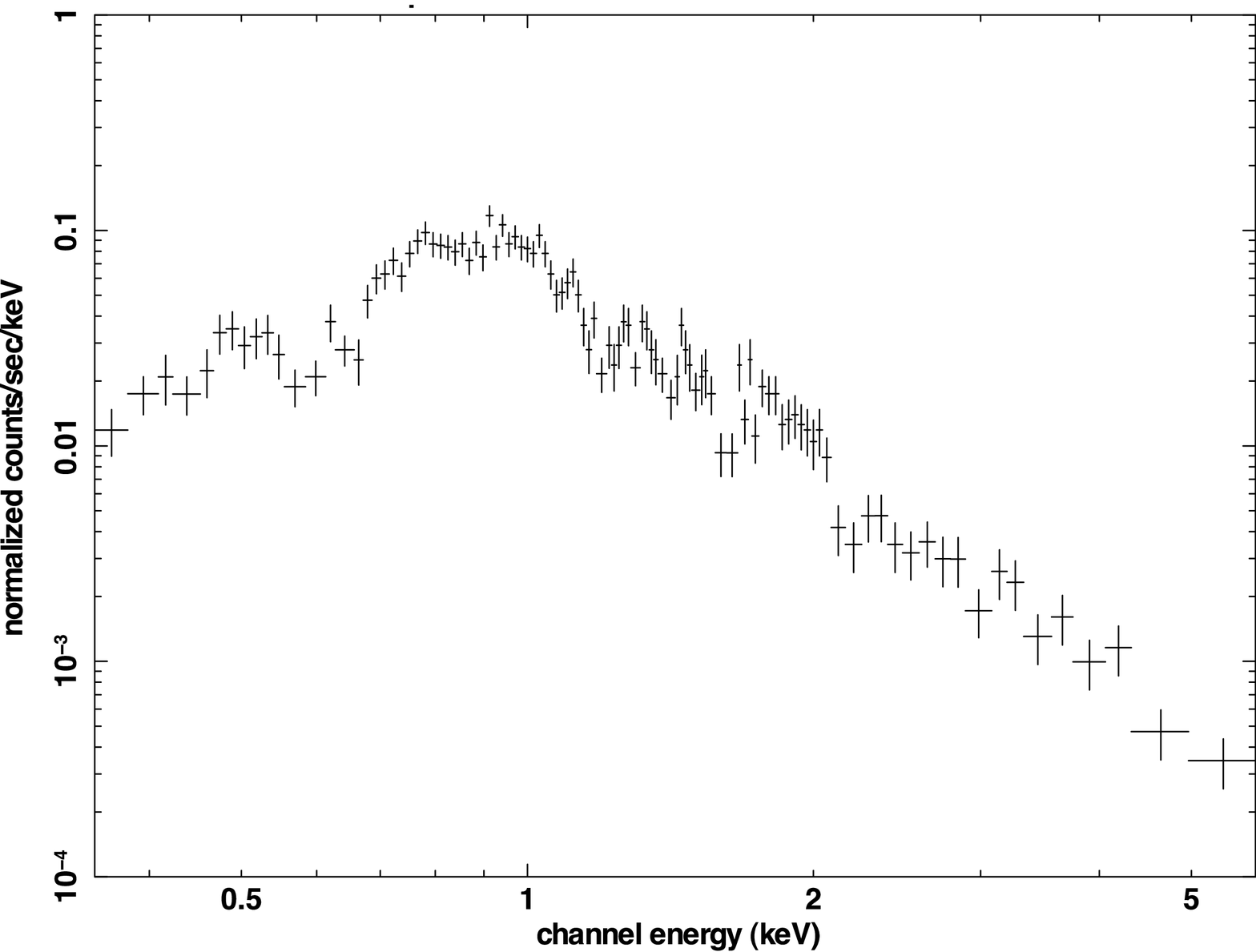}}}
\resizebox{0.4\hsize}{!}{\rotatebox{0}{\includegraphics{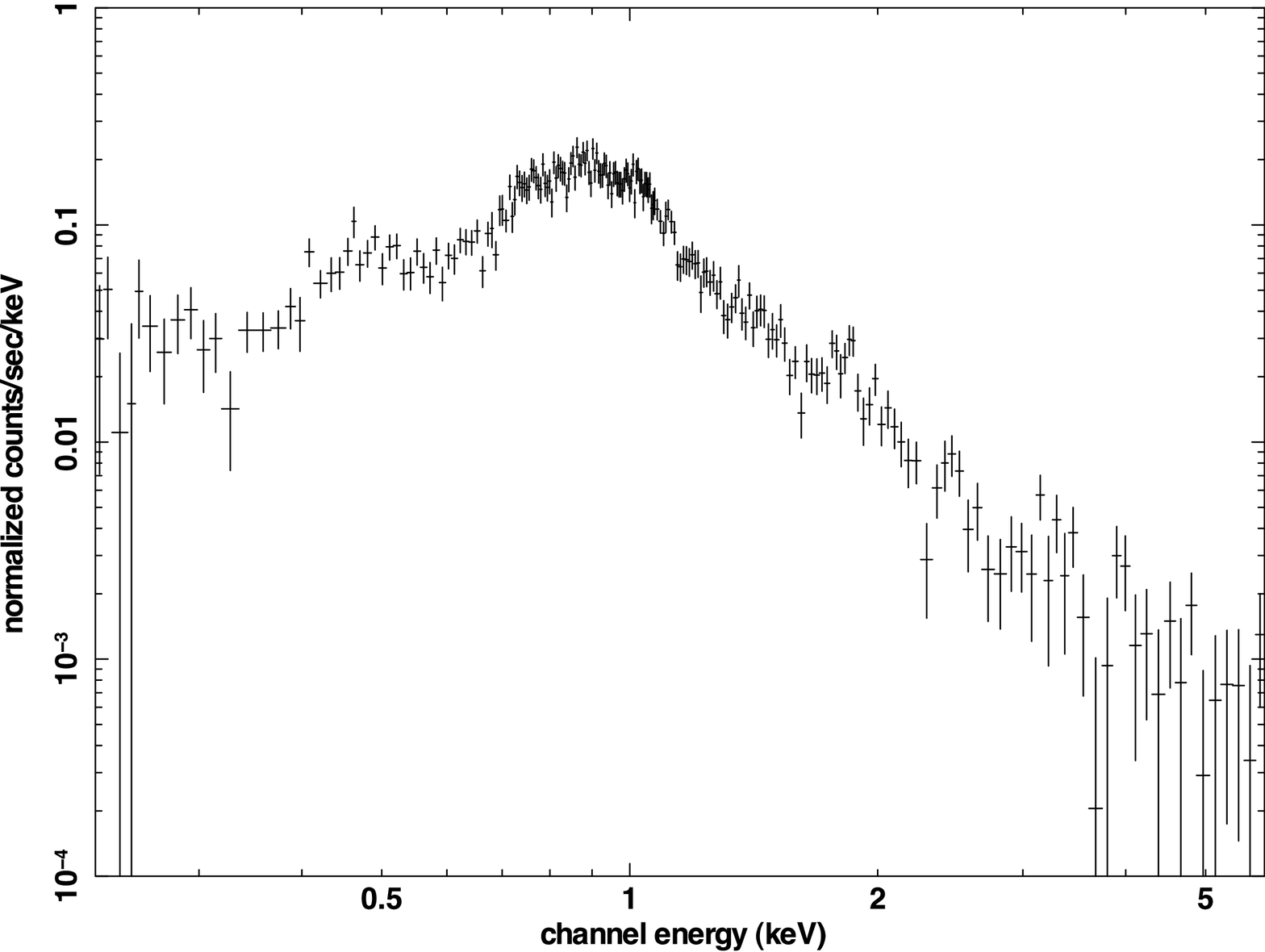}}}
\end{center}
\caption{The Chandra dataset  (left) at t=2594 days since explosion and the XMM dataset (right) at t=2943 days since explosion.} 
\label{fig:obs-data}
\end{figure*}

The parent galaxy of SN 1993J, M81, was observed
 with a 50ks exposure on May 7, 2000 with the Chandra Advanced
CCD Imaging Spectrometer (ACIS) spectroscopy array operating
in imaging mode \citep{Swartz03}.
We refer to \citet{Swartz03} for the technical details of the
observations.
We extracted the data from the ChandraXO archive and 
analyzed this using standard CIAO analysis threads and XSPEC \citep{arn96}.
In order to check for any X-ray flaring events,
we selected a background region of
size 90$''$, carefully excluding point sources.
We used a radius of 7$''$ centered at the 
SN 1993J position to extract the SN events;
and an annulus of
inner radius 7$''$ and outer radius 18$''$ centered at the
supernova position
to obtain the  background counts.
We refer to \citet{Swartz03} and \citet{Chandra05} for the data
analysis procedure.
We binned the data in 15 cts per channel for the spectral analysis. The observed spectrum is shown in Fig~\ref{fig:obs-data}. While analysing the supernova spectrum with XSPEC, we ignored the 
counts below 0.2 keV and above 6.0 keV, since there was
almost no flux in these energy ranges.

\subsection{The XMM dataset}

XMM-Newton observed SN1993J
on April 22/23, 2001, for a total
of 132 ksec. The 90 ks data was with the PN camera in the small
window mode and 83 ks with the MOS2 camera in image mode. 
The medium filter was applied both in the PN and the
MOS2 instrument setup during the observation 
\citep{ZimAsch03}.
Due to the presence of the  variable particle background in parts of the
observing period, we flagged the flaring events from the light curve.
In the PN observations about
half of the observations were affected by bursts in the particle
background. We extracted the spectrum and created response matrices using 
XMM-SAS.
The spectrum was
binned as group of 50 bins. The observed spectrum is shown in Fig.~\ref{fig:obs-data}. Spectral fits were performed using only the PN camera data
offering the best statistics.

\section{Modeling}
In general, the X-ray flux is a combination of the flux from the forward and reverse shocks.
As we will show below, in most cases the reverse shock is likely to be radiative. For these cases we have made a self-consistent model in which hydrodynamic calculations for a stationary flow are combined with time-dependent ionization balance and multi-level atoms, including optical depth effects  \citepalias{Nymark06}. This means that the emission in the cooling flow behind the reverse shock is followed from the shock to the cool shell behind the shock, giving correct weight to the emission from each zone. The parameters of the shock calculations are the reverse shock temperature, the density and the chemical composition, and the output is the emitted spectrum. For the cases where the cooling time is long we assume a one-temperature spectrum.

Ionization rates are from \citet{ArnRoth} and \citet{ArnRay}, while recombination rates are from \citet{Nah96,Nah99}, \citet{NahPr97}, and \citet{Zat03,Zat04a,Zat04b}. Most of the collisional and radiative data are from the Chianti database, version 4.2~\citep{Dere97, Young03}\footnote{CHIANTI is a collaborative project involving the NRL (USA), RAL (UK), and the following universities: College London (UK), Cambridge (UK), George Mason (USA), and Florence (Italy).}. For further details we refer to \citetalias{Nymark06}.

\section{Structure of the ejecta}

The properties of the circumstellar interaction depend on the density structure in the ejecta. Explosion models for red supergiants indicate that the density distribution in the ejecta can be described by a power law, $\rho \propto v^{-\eta }$. For a polytrope of index $n=3/2$, typical of red supergiants, \citet{MatznerMcKee99} have shown that $\eta \approx 10$ in the outer layers. Deviations from a simple polytropic progenitor may, however, result in a more complicated structure. As an example, the model 4H47, specifically calculated for SN~1993J \citep{Shigeyama94, NomHash88}, shows a structure with a steep outer part and a flatter inner profile, and then an additional steepening further in, as well as considerable structure in the inner parts (see their Fig.~3). This profile is, however, also a simplification, since \citet{Iwamoto97} find that especially the inner, steep region is Rayleigh-Taylor unstable.

As the reverse shock moves inward in the ejecta, it passes slowly through the steep outer gradient, while moving more rapidly in the flatter part before eventually reaching the second steep gradient. When the gradient changes, the spectrum is also expected to change as the reverse shock slows down (steep part) or speeds up (flat part). There is evidence that this has happened in SN~1993J \citep{Dwarkadas05}. 

The shock radius depends on time as $R_{\mathrm{s}}\propto t^{m}$, where the expansion parameter $m=(\eta -3)/(\eta -2)$. Numerical calculations by \citet{Dwarkadas05} based on the 4H47 model indicate that a change in the expansion parameter should occur at approximately day 2300. Such a change was indeed observed by \citet{Bartel02} in the radio at $\sim 1400$ days. For the cases discussed in the present work it is worth noting the difference in the X-ray spectra from Chandra \citep{Swartz03} on day 2594 and the one from XMM \citep{ZimAsch03} a year later. Although both these observations were done somewhat later than the expected change in the spectrum, the observed difference could be related to a change in the expansion parameter.

\begin{table*}[h*]
\caption{Fractional abundances by number and molecular weights for the compositions discussed here.}
\label{tab:abund}
\begin{center}
\begin{tabular}{lcccccccc} \hline \hline 
 & &  & 4H47 &  &  &s15s7b &  & \ \\ \hline \\
\phantom & Solar & H/He zone & He/N zone & C/O zone &H/He zone & He/N zone & C/O zone & SN 1987A-like \ \\ \hline \\
H & $9.10\times 10^{-1}$& $7.41\times 10^{-1}$ &  --- & ---  & $8.87\times 10^{-1}$ & --- & --- & $7.99\times10^{-1}$\ \\ 
He &$ 8.89\times 10^{-2}$& $2.57\times 10^{-1}$ &$ 9.95\times10^{-1}$& $9.66\times 10^{-2}$& $1.11\times 10^{-1}$ &$9.95\times 10^{-1}$& --- & $2.0\times 10^{-1}$ \\ 
C  &$3.79\times 10^{-4}$& $3.01\times 10^{-5}$ &$6.77\times 10^{-5}$& $4.07\times 10^{-1}$& $2.23\times 10^{-4}$ &$8.76\times 10^{-5}$&$2.36\times 10^{-1}$& $1.03\times 10^{-4}$\\ 
N & $7.92\times 10^{-5}$& $1.29\times 10^{-3}$ & $2.92\times 10^{-3}$& --- & $3.49\times 10^{-4}$& $3.64\times 10^{-3}$  &-- & $5.15\times 10^{-4}$\\ 
O & $6.29\times 10^{-4}$& $3.38\times 10^{-5}$ & $5.08\times 10^{-5}$&$4.80\times 10^{-1} $& $6.99\times 10^{-4}$ &$6.49\times 10^{-5}$&$7.43\times 10^{-1}$& $4.69\times 10^{-4}$\\ 
Ne & $8.89\times 10^{-5}$& $7.86\times 10^{-4}$ & $1.77\times 10^{-3}$&$1.23\times 10^{-2} $& $1.19\times 10^{-4}$ &$3.56\times 10^{-4}$&$1.93\times 10^{-2}$& $8.89\times 10^{-5}$\\ 
Mg &$ 3.62\times 10^{-5}$ & $3.57\times 10^{-5}$ &$ 8.03\times 10^{-5}$&$ 3.21\times 10^{-3}$& $3.92\times 10^{-5}$ &$1.17\times 10^{-4}$&$7.69\times 10^{-4}$& $3.63\times 10^{-5}$\\ 
Si &$3.46\times 10^{-5}$& $4.42\times 10^{-5}$ &$9.94\times 10^{-5}$ & $3.25\times 10^{-4}$& $3.73\times 10^{-5}$ &$1.11\times 10^{-4}$&$4.16\times 10^{-4}$& $3.45\times 10^{-5}$\\ 
S & $1.69\times 10^{-5}$ & $2.13\times 10^{-5}$ & $4.78\times 10^{-5}$& $1.56\times 10^{-4}$& $1.81\times 10^{-5}$ &$5.40\times 10^{-5}$&$2.02\times 10^{-4}$& $1.69\times 10^{-5}$\\ 
Ar  & $3.62\times 10^{-6}$& $9.30\times 10^{-6}$ & $2.08\times 10^{-5}$& $6.13\times 10^{-5}$& $3.63\times 10^{-6}$ &$1.08\times 10^{-5}$& $4.05\times 10^{-5}$& $3.62\times 10^{-6}$\\ 
Ca  &$2.13\times 10^{-6} $& $2.63\times 10^{-6}$ &$5.93\times 10^{-6} $& $1.94\times 10^{-5}$& $2.23\times 10^{-6}$ &$6.66\times 10^{-6}$&$2.49\times 10^{-5}$& $2.13\times 10^{-6}$\\ 
Fe & $3.08\times 10^{-5}$& $5.30\times 10^{-5}$ & $1.19\times 10^{-4}$&$3.90\times 10^{-4}$ & $3.28\times 10^{-5}$ &$9.79\times 10^{-5}$&$3.66\times 10^{-4}$& $3.08\times 10^{-5}$\\
Ni & $ 1.62\times 10^{-6}$& $2.78\times 10^{-6}$ & $ 2.22\times 10^{-6}$&$2.05\times 10^{-5}$& $1.72\times 10^{-6}$ &$5.14\times 10^{-6}$&$1.92\times 10^{-5}$& $1.62\times 10^{-6}$\\ 
$\mu _{\mathrm{A}}$& 1.29 \phantom{ } \phantom{ } \phantom{ } \phantom{ } \phantom{ }&  1.81 \phantom{ } \phantom{ } \phantom{ } \phantom{ } \phantom{ }& 4.07 \phantom{ } \phantom{ } \phantom{ } \phantom{ } \phantom{ } &11.37  \phantom{ }\phantom{ } \phantom{ } \phantom{ } \phantom{ } \phantom{ } &  1.36 \phantom{ } \phantom{ } \phantom{ } \phantom{ } \phantom{ } & 4.06 \phantom{ } \phantom{ } \phantom{ } \phantom{ } \phantom{ } & 15.20  \phantom{ } \phantom{ } \phantom{ } \phantom{ } \phantom{ } \phantom{ }&  1.63 \phantom{ } \phantom{ } \phantom{ } \phantom{ } \phantom{ } \\
$\mu $ & 0.61\phantom{ }  \phantom{ } \phantom{ } \phantom{ } \phantom{ } &  0.80  \phantom{ } \phantom{ } \phantom{ } \phantom{ } \phantom{ }& 1.34  \phantom{ } \phantom{ } \phantom{ } \phantom{ } \phantom{ }& 1.70 \phantom{ } \phantom{ } \phantom{ } \phantom{ } \phantom{ } &   0.64  \phantom{ } \phantom{ } \phantom{ } \phantom{ } \phantom{ }&  1.34  \phantom{ } \phantom{ } \phantom{ } \phantom{ } \phantom{ }& 1.77  \phantom{ } \phantom{ } \phantom{ } \phantom{ } \phantom{ }&   0.74  \phantom{ } \phantom{ } \phantom{ } \phantom{ } \phantom{ } \\
\hline \\
\end{tabular}
\end{center}
\end{table*}

Models of the light curve and spectrum of SN~1993J strongly indicate that less than $0.5~M_{\odot}$ of the hydrogen envelope was left at the time of the explosion \citep[\eg ][]{Woosley94, Nomoto93, Ray93, Houck96}. Further, explosion models like the 4H47 model by \citet{NomHash88} and \citet{Shigeyama94} predict that the outer parts of the ejecta should contain nearly equal mass fractions of H and He as a result of CNO processing. 

Assuming a constant density gradient in the ejecta and a steady wind, we can estimate the swept up mass and therefore what the composition at the reverse shock should be. If the density of the CSM is given by 

\begin{equation}
\label{eq:windDens}
\rho _{\mathrm{w}}=\frac{\dot{M} }{4\pi R_{\mathrm{s}}^{2}v_{\mathrm{w}}}, 
\end{equation}

\noindent
the mass swept up by the reverse shock is

\begin{equation}
M_{\mathrm{rev}}=\frac{\eta -4}{2}\frac{\dot M}{v_{\mathrm{w}}}R_{\mathrm{s}},
\end{equation}

\noindent
where $v_{\mathrm{w}}$ is the wind velocity \citep{Chevalier1982b}. With $R_{\mathrm{s}}=V_{\mathrm{ej}}t$, we get

\begin{eqnarray}
M_{\mathrm{rev}}&=&1.3\times 10^{-2}(\eta -4)\left(\frac{V_{\mathrm{ej}}}{10^4\ \mathrm{km\ s^{-1}}}\right)\\
& \times &\left(\frac{\tilde{A_{*}}}{4}\right)\left(\frac{t_{\mathrm{d}}}{1000\ \mathrm{days}}\right)\ \ \mathrm{M_{\odot}}, \nonumber
\end{eqnarray}

\noindent
where $V_{\mathrm{ej}}$ is the ejecta velocity and $t_{\mathrm{d}}$ the time since explosion in days. For SN~1993J $\tilde{A_{*}}\equiv (\dot{M}/10^{-5}\ \mathrm{\Msun \ yr^{-1}})/(v_{\mathrm{w}}/10 \  \mathrm{km\ s^{-1}})\approx 4$ \citepalias{FLC96}. For the Chandra observations we take $t_{\mathrm{d}}=2600$ and $V_{4}=V_{\mathrm{ej}}/10^4\mathrm{km\ s^{-1}}\approx 1$, which is consistent with the expansion velocity derived from VLBI observations at this time \citep{Bartel02, Bartel07}.

Assuming an ejecta density gradient $\eta =25$ in the outer zones, and $\eta =5$ in the shallow middle zone, as indicated by the 4H47 model by \citet{Shigeyama94}, we find the swept up mass at 2600 days to be $\sim 0.38\ \Msun$. 
The hydrogen-rich envelope in the 4H47 model has a mass of $0.47\ \Msun $. Inside this envelope the outer part of the He zone is N-rich, while the inner part is C-rich. The He/N zone contains $0.3\ \Msun $, and the He/C zone contains $1.5\ \Msun $. 

The estimate above shows that the reverse shock may be close to the border between the H-rich envelope and the He-rich zone. However, this estimate depends on the mass assumed for the hydrogen envelope. With a lower hydrogen mass, as in the s15s7b model by \citet{Woosley94}, where the hydrogen envelope is only $0.2\ \Msun$, it is therefore quite possible that the reverse shock is somewhere in the He-rich zone. Nevertheless, we cannot exclude the possibility that clumps of metal-rich material may have reached the outer parts by mixing, as indicated by the simulations by \citet{Iwamoto97}. In Sect.~\ref{sect:results} we therefore also discuss the spectrum resulting from metal-rich compositions. 

\citet{Swartz03} found that the Chandra spectrum of SN~1993J was best fit with a composition dominated by H, and where N, Mg, Si, and Fe were enhanced with respect to solar abundances, while He, C, O and Ne were absent. These abundances are, however, not consistent with the calculated models for the ejecta. 
In the analysis by \citet{ZimAsch03} the XMM observations were fit with a number of different models, all with widely differing elemental abundances, but these abundances were very poorly constrained.

\section{Radiative shocks}
The structure of the reverse shock was discussed in \citetalias{Nymark06}, where we showed that this can remain radiative for several years, and longer. The emission and cooling depend on the reverse shock temperature, the pre-shock density and the composition. These are the fundamental parameters, but each of these are related to the shock velocity, $V_{\mathrm{s}}$, mass loss parameter, $A_*$, ejecta density gradient, $\eta $, and the time since explosion, $t$, by the relations discussed in NFK06. For clarity we here repeat the relevant relations.   

An estimate of whether the shock is radiative can be found by comparing the expansion time with the cooling timescale. In \citetalias{Nymark06} we showed that the ratio between these timescales can be written as

\begin{equation}
\frac{t_{\mathrm{cool}}}{t} =  \frac{6.7\times 10^{-22}\mu _{\mathrm{A}}^2\mu V_{4}^4 t_{\mathrm{d}}}{(\mu _{\mathrm{A}}-\mu )(\eta -2)^2(\eta -3)(\eta -4)\Lambda (T_{\mathrm{rev}}) \tilde{A}_{*}},
\label{eq:tcool}
\end{equation}

\noindent
where $\mu $ is the mean mass per particle and $\mu _{\mathrm{A}}$ is the mean atomic weight. The cooling function, $\Lambda (T_{\mathrm{rev}})$ is discussed in \citetalias{Nymark06}. The shock is radiative if $t_{\mathrm{cool}} \la t$. If the interaction region is thin compared with the size of the supernova, as will be the case if the shock is radiative, the thin shell approximation of \citet{Chevalier1982b} applies. In this case the temperature of the reverse shock is given by

\begin{equation}
T_{\mathrm{rev}}=\frac{2.27\times 10^9\mu }{(\eta -2)^2}V_{4}^2.
\end{equation}

\noindent
Using this in Eq.~(\ref{eq:tcool}) and requiring that $t_{\mathrm{cool}}/t\la 1.0$ we find that the shock is radiative if the reverse shock temperature fulfills the requirement

\begin{eqnarray}
T_{\mathrm{rev}}&\la &\frac{8.8\times 10^{19}}{\mu _{\mathrm{A}}(\eta -2)}\\
&\times &\left (\mu (\mu_{\mathrm{A}}-\mu)(\eta -3)(\eta -4)\Lambda (T)\tilde{A_{*}}t_{\mathrm{d}}^{-1}\right )^{\frac{1}{2}}.
\end{eqnarray}

\begin{figure*}[ht]
\begin{center}
\resizebox{0.6\hsize}{!}{\rotatebox{0}{\includegraphics{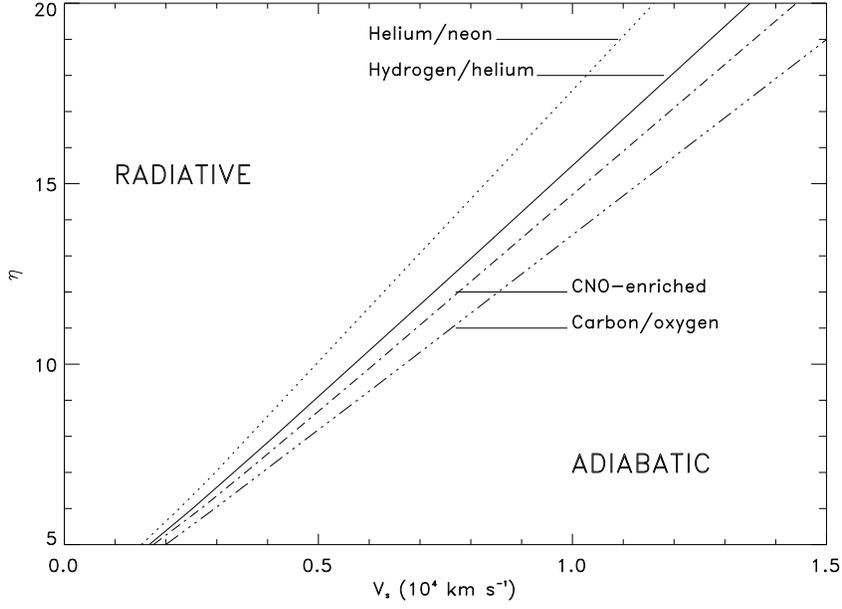}}}
\end{center}
\caption{The lines represent the border between the radiative and adiabatic regimes for each composition. In each case the region above the line corresponds to the reverse shock being radiative, while values of $\eta$ and $V_{\mathrm{s}}$ below the line lead to adiabatic shocks.} 
\label{fig:limVel-eta}
\end{figure*}

Table~\ref{tab:limTemp} shows the limiting temperatures, $T_\mathrm{rad}$, and expected shock temperatures for a number of compositions and density gradients for the case of SN~1993J, with $\tilde{A_{*}}=4$ and $t_{\mathrm{d}}=2600$~days. $T_\mathrm{rad}$ is defined as the limiting temperature corresponding to $t_{\mathrm{cool}}=t$, such that if $T_{\mathrm{rev}} < T_{\mathrm{rad}}$ the shock will be radiative while $T_{\mathrm{rev}} > T_{\mathrm{rad}}$ gives an adiabatic shock. These regimes are also illustrated in Fig. \ref{fig:limVel-eta} in terms of the ejecta density gradient and the shock velocity. The lines in Fig. \ref{fig:limVel-eta} represent the border between the adiabatic and radiative regimes for a few different compositions. If the ejecta density gradient and the shock velocity are in the region above the line corresponding to the relevant composition, the shock is radiative. This means that for a given shock velocity, the shock is more likely to be radiative the higher the density gradient is. We emphasize that the values of $\eta $ and $V_{\mathrm{s}}$ corresponding to radiative and adiabatic shocks shown in Fig.~\ref{fig:limVel-eta} are a function of the time since the explosion, the mass loss rate and the cooling rate, and are therefore only valid for the specific case discussed here.

\begin{table*}[h*]
\caption{Reverse shock temperature, $kT_{\mathrm{rev}}$ and limiting temperature, $kT_{\mathrm{rad}}$ (both in keV) for a radiative shock,  for SN~1993J at 2600 days for $\dot{M}=4.0\times 10^{-5} \mathrm{M_{\odot}\ yr^{-1}}$.} 
\label{tab:limTemp}
\begin{center}
\begin{tabular}{l|cc|cc|cc} \hline \hline 
 \multicolumn{1}{c|}{$\eta$}&  \multicolumn{2}{c|}{7}   &  \multicolumn{2}{c|}{12}    &  \multicolumn{2}{c}{20} \\ \hline 
Composition  &$kT_{\mathrm{rad}}$& $kT_{\mathrm{rev}}$&$kT_{\mathrm{rad}}$& $kT_{\mathrm{rev}}$&$kT_{\mathrm{rad}}$& $kT_{\mathrm{rev}}$\\ \hline
Solar   & 0.51 & 4.77 & 0.63 & 1.19 & 0.68 & 0.37   \\ 
H/He (4H47)& 0.69 & 6.30 & 0.82 & 1.57 & 0.88 & 0.48 \\ 
He/N  (4H47)& 0.93 & 10.5& 1.04 & 2.60 & 1.09 & 0.81 \\ 
C/O (4H47) &  2.02 & 13.5 & 2.43 & 3.39 & 2.61 & 1.04 \\
H/He (s15s7b2f) & 0.53 & 5.01 & 0.66 & 1.25  & 0.71 & 0.39 \\ 
He/N (s15s7b2f)  & 0.86 & 10.4 & 0.99 & 2.60 & 1.04 & 0.80  \\ 
C/O (s15s7b2f)  & 2.32 & 13.9 & 2.80 & 3.46 & 3.00 & 1.07\\ 
SN~1987A-like& 0.73 & 5.70& 0.87 & 1.40  & 0.92 & 0.44 \\
\hline 
\end{tabular}
\end{center}
\end{table*}

\citet{Swartz03} fit the SN 1993J data with two low temperature components ($kT=0.35$ keV and
$kT=1.0$ keV) and one high temperature component ($kT=6.0$ keV). 
From Table~\ref{tab:limTemp} we find that the $kT_{\mathrm{rev}}=0.35$~keV  shock is likely to be radiative, and, depending on the composition, also the $kT_{\mathrm{rev}}=1.0$~keV component, while a $kT_{\mathrm{rev}}=6.0$~keV schock is probably adiabatic. This means that the assumption of a single-temperature shock is not realistic for the two low-temperature components, and their analysis is therefore not consistent. 

It is possible that the high energy component comes from the hot outer shock, while the $kT_{\mathrm{rev}}=1.0$~keV component comes from a radiative reverse shock. In this case, the $kT_{\mathrm{rev}}=0.35$~keV component could be representative of the cooling part downstream from the reverse shock. However, as we discuss below, in Sect.~\ref{sect:csEmission}, the temperature of the outer shock is unlikely to be as low as 6.0~keV. Another possibilty is the presence of several shocks in  the ejecta, for example if the ejecta is clumpy. In that case all three components could come from the reverse shock, but corresponding to different angles with respect to the clumpy ejecta. This alternative is  discussed in more detail in Sect.~\ref{sect:oblique}.

\section{Emission from the circumstellar shock}
\label{sect:csEmission}
At late times the reverse shock is expected to dominate the X-ray emission. There may, however, still be a contribution from the outer shock, in particular at energies $\ga 5$~keV. The high temperature component found in the fits to the Chandra and XMM observations of SN~1993J was attributed to the circumstellar shock, which is expected to have a considerably higher temperature than the reverse shock. It is surprising, though, that this temperature is as low as $\sim 6.0$~keV. The shocked CSM has an ion temperature of \citepalias{FLC96}

\begin{equation}
\label{eq:Tcs}
T_{\mathrm{cs}}=2.27\times10^9\mu \frac{(\eta -3)^2}{(\eta -2)^2}V_{4}^{2}\ \ \ \mathrm{K},
\end{equation}

\noindent
where  $\mu =0.61$ for a solar composition, and given in Table~\ref{tab:abund} for other compositions. For $V_{4}=1$ and $\eta =12$, $kT_{\mathrm{cs}}\ \gsim \ 100$~keV. Allowing for the possibility of a lower ejecta density gradient or shock velocity, this value may be reduced somewhat, but it is difficult to get a temperature lower than $\sim 40$~keV if the ion temperature is equal to the electron temperature. 

However, if collisionless heating is inefficient, the electron temperature can be considerably lower than the ion temperature \citepalias[e.g.,][]{FLC96}.  A lower limit to the electron temperature may be found 
from the timescale for electron-ion equilibration 

\begin{eqnarray}
t_{\mathrm{e-i}}&\approx &29 \left(\frac{T_{\mathrm{e}}}{10^{9}\
\mathrm{K}}\right)^{3/2}\left(\frac{n_{\mathrm{e}}}{10^8\
\mathrm{cm^{-3}}}\right)^{-1}\nonumber \\
& \approx & 0.21\left(\frac{T_{\mathrm{e}}}{10^{9}\
\mathrm{K}}\right)^{3/2}\tilde{A_{*}}^{-1}V_{4}^{2}t_{\mathrm{d}}^{2}\ \ \rm{days}.
\end{eqnarray} 

Collisionless heating by plasma instabilities may, however, increase the electron temperature. There is, however, evidence from observations of galactic supernova remnants that collisionless heating is only effective for low Mach number shocks \citep{Ghavamian07, Vink03}. Because of the much higher densities at these comparatively early stages, it is, however, not clear how these results apply to our cases.

Equating $t_{\mathrm{e-i}}$ to the expansion time gives

\begin{equation}
\label{eq:elTemp-lim}
kT_{\mathrm{e}} \ga 7.4 V_{4}^{-4/3}\left ( \frac{\tilde{A_{*}}}{4}\right)^{2/3}\left(\frac{t_{\mathrm{d}}}{1000\ \mathrm{days}} \right )^{-2/3} \ \ \ \mathrm{keV.}
\end{equation}

\noindent
An electron temperature as low as 6.0~keV, as found by \citet{Swartz03}, is therefore in principle possible.  Also cosmic ray dominated shocks can give a lower temperature and higher compression rate \citep[e.g.,][]{Decourchelle00, Ellison04}, which could contribute to lowering the temperature of the forward shock. For efficient cosmic ray acceleration, with an injection efficiency (fraction of protons with superthermal energies) of $\eta=10^{-2}$ at the outer shock the models of \citet{Decourchelle00} give temperatures of a few keV, while the resulting high compression can lead to relative thickness of the outer shock (as defined below) of $f\approx 0.04$. Also this effect can therefore  give a shock temperature of 6.0~keV. 

However, a second requirement is that this component should also produce the observed luminosity of $1.6\times 10^{38} \ \mathrm{ergs\ s^{-1}}$ in the range 2.4-8.0~keV. The luminosity from the circumstellar shock is given by

\begin{eqnarray}
\label{eq:lumCS}
L_{\lambda,\mathrm{cs}}& \approx &2.54\times 10^{37} \frac{\mu _{\mathrm{A}}-\mu}{\mu \mu _{\mathrm{A}}} \tilde{A}_{*}^{2}V_{4}^{-1}\left (\frac{f}{0.2}\right )^{-1} \nonumber \\
& &\times T_8^{-0.236}\lambda ^{-2} e^{-1.44/\lambda T_8} \nonumber \\
& & \times \left
(\frac{t_{\mathrm{d}}}{1000\  {\mathrm{days}}}\right )^{-1}\ \ \mathrm{erg\
s^{-1}\AA ^{-1}},
\end{eqnarray}

\noindent
using a Gaunt factor $g_{\mathrm{ff}}\sim 1.87T_{8}^{-0.264}$~\citepalias{FLC96}. In this expression $f$ is the relative thickness of the adiabatic outer shock, defined by $f=(R_{\mathrm{cs}}-R_{\mathrm{cd}})/R_{\mathrm{cd}}$, where $R_{\mathrm{cs}}$ is the outer shock radius and $R_{\mathrm{cd}}$ is the radius of the contact discontinuity. For $\eta =7 - 20$,  $f\approx 0.2-0.3$~\citepalias{CF94}.
Inserting the appropriate parameters, we find that the expected luminosity in the range 2.4--8.0~keV from a circumstellar component with $kT\sim 6.0$~keV is $L_{\mathrm{cs}}\sim 2.0\times 10^{37} \ \mathrm{ergs\ s^{-1}}$ at 2600 days, \ie only about one tenth of the observed luminosity. A higher compression, as resulting from cosmic ray acceleration, can lead to higher luminosity. However, even the most efficient model in \citet{Decourchelle00}, with $\eta =10^{-2}$, which gives a compression of  $f\approx 0.04$, will  for a temperature of 6.0~keV lead to an increase of the luminosity by only a factor of five, which is not enough to explain the observed luminosity. We conclude that it is  unlikely that the circumstellar shock is responsible for the 6.0~keV component. This is in agreement with \citetalias{FLC96} who find that the luminosity of the forward shock after $t \sim 1000$ days is orders of magnitude lower than that of the reverse shock.

\section{Oblique shocks}
\label{sect:oblique}
If either the CSM or the ejecta is clumpy, parts of the reverse shock will hit these clumps obliquely. The transmitted shocks will then have a lower velocity than the part of the reverse shock which hits the clumps head-on, and also the shock temperature will be lower. This scenario could result in a considerable range of shock velocities contributing to the emission. In particular, even if the head-on shock is adiabatic for the parameters chosen (\ie a given reverse shock velocity and ejecta density), the tangential shocks may be radiative.

This is similar to the situation found for  the ring collision in SN~1987A, where \citet{Zhekov06} find that a range of shock velocities are necessary to explain the observed spectrum. The presence of both hard X-rays and optical emission from the shocked CSM clearly shows the importance of geometric and complex dynamics of a clumpy CSM. The hydrodynamical simulations by \citet{Iwamoto97}, as well as \citet{Kifonidis06}, however, show that also the ejecta is subject to Rayleigh-Taylor instabilities in connection to the passage of the shock. The same situation as in the CSM case with a range of velocities for the reverse shock will then arise. With this background, we therefore attempt to fit the spectra with two or more shocks, possibly with a mix of adiabatic and radiative shocks. The results of this fitting are discussed in the next section.

\section{Model fitting of the X-ray spectra}
\label{sect:results}
In this section we discuss the modeling of the spectra along the lines of the previous discussion. First we emphasize that we do not aim at perfect statistical fits, keeping all parameters, like temperature, abundances and absorbing column density, free. Instead, our aim is to investigate how well -- or how badly -- the observations are reproduced by different realistic explosion models. 
Therefore we focus on the important spectral features, and investigate how well each of them is reproduced by each model. In this way we hope to distinguish between different models for the composition, and therefore the nature of the progenitor. We believe this to be more physically illuminating than the more commonly used method of varying all parameters until a best fit has been achieved. In particular, with data of low spectral resolution and moderate S/N this can easily result in doubtful abundances. Different sets of abundances may result in similar values of $\chi^2$, and the one with the lowest value may have little significance compared to others with substantially different abundances. Since we do not allow all parameters to vary freely, the  $\chi^2$ statistic is not really appropriate. One approach which is used in similar circumstances is to investigate quantities which can be determined unambiguously in the data and the model \citep[e.g.,][]{Badenes08}. In the present case, however, the quality of the data does not allow any this to be determined unambiguously. Therefore we choose instead to focus on specific spectral features, and use the  $\chi^2$ statistic as a guideline to distinguish between a few specific models. This is similar to the approach taken by \citet{Rakowski06}.

We have chosen to relate our models to the ejecta composition found from two explosion models for SN~1993J, the 4H47 model \citep{NomHash88, Shigeyama94} and the s15s7b model \citep{Woosley94}. From each of these we have selected a number of compositions, corresponding to the different burning zones in the ejecta (Table~\ref{tab:abund}).  

We also discuss a CNO-enriched composition resembling the one in the circumstellar ring of SN~1987A, but with initially three times higher metallicity, to account for the low metallicity of the LMC compared to the more normal metallicity in M81 and NGC~3877. This composition  has been determined in the following way: We took H/He=0.25 by number, as appropriate for the circumstellar ring of SN~1987A. The sum of C, N and O abundances was taken to be equal to the solar value, while the ratios were the same as in the ring of SN~1987A, \ie N/C=5 and N/O=1.1 \citep{LF96}. For all other elements we use solar abundances. 
We note that the optical observations of SN~1993J indicate a high N/C ratio \citep{Fransson05}. 

We also note that due to the presence of the cooling shell, part of the flux density from the reverse shock will be absorbed \citepalias[\eg][]{CF94}. This absorption will come in addition to the Galactic absorption column density. However, unlike the Galactic absorption, which approximately should have solar abundances, this extra absorption should normally have 
the abundances of the burning zone close to the shock. 
We therefore first determine the column density for a nearby source in the field of view of the SN, and freeze the Galactic column density to that value in our fitting. The extra absorption contribution
will then solely come from the cooling shock. 

 For each composition we attempt to fit the spectrum with one reverse shock temperature, which we allow to vary between the maximum reverse shock temperatures corresponding to ejecta density gradients in the range $\eta=7$ to $20$. The corresponding reverse shock temperature, $T_{\mathrm{rev}}$, for each composition is given in Table~\ref{tab:limTemp}, together with the limiting temperature for a radiative shock, $T_{\mathrm{rad}}$. If the best fit reverse shock temperature exceeds $T_{\mathrm{rad}}$, we take an 
adiabatic component with
a temperature that is allowed to vary between $T_{\mathrm{rad}}$ and $T_{\mathrm{rev}}$. 

We also tested adding a hot, single-temperature component with solar abundances, representing the outer shock. For this component we used the XSPEC model ``Vmekal'', which gives the spectrum from a hot, diffuse gas with a single temperature. This is basically a free-free spectrum for $kT_{\mathrm{cs}}\ga 10$~keV.

\section{Results}
The XMM-Newton spectrum of SN~1993J  (Fig.~\ref{fig:obs-data}, right hand panel) is dominated by the Fe~L emission at 0.7--1.0 keV, which is blended with strong \ion{Ne}{ix--x} emission. A lower peak is evident at 0.4--0.5 keV, most likely caused by strong \ion{N}{vii} lines. Several emission lines are present above $\sim 1.0$~keV, in particular from \ion{Mg}{xi-xii} and \ion{Si}{xiii-xiv}, with the \ion{Mg}{xi} and \ion{Si}{xiii} lines being the stronger, as well as \ion{S}{xvi}. The absence of O emission  at 0.6-0.8 keV is striking, as the \ion{O}{viii} lines at 0.65 and 0.775 keV are expected to be strong over a wide range of temperatures. 

The Chandra spectrum  (Fig.~\ref{fig:obs-data}, left hand panel) shows much the same features as the XMM spectrum, but with a more pronounced double appearance of the Fe~L peak, with one peak at $\sim 0.8$ keV and a second peak at $\sim 0.9$ keV. Because the S/N is considerably lower for the Chandra spectrum, we will in the following concentrate on the XMM spectrum.

We will now systematically investigate to what extent the abundance models in Table~\ref{tab:abund} reproduce these observations. We start with a solar composition, and then go on to more specialized models for different nuclear burning zones.  

In the case of SN 1993J \citet{Swartz03} found the best fit column density for the M81 nucleus to be $4.4\times 10^{20}$ cm$^{-2}$. \citet{Dickey90} found that the Galactic absorption column density in that direction is $4.17\times 10^{20}$ cm$^{-2}$, which is well in agreement with
that of the best fit column density. Hence we froze the Galactic column density with
solar composition to this value.

When fitting the spectrum with one radiative shock, we found that the best fit temperature exceeded $T_{\mathrm{rad}}$ in all models. An adiabatic shock with a higher temperature, gave a better fit, but in most cases the best fit was found for a combination of an adiabatic shock and a radiative shock at half the temperature of the adiabatic shock. 

Adding a hot component from the circumstellar shock did not improve the fits, and the best fit temperature for this component was $\la 5$~keV in all cases. As discussed in Sect.~\ref{sect:csEmission}, the forward shock is unlikely to have such a low temperature. We therefore conclude that there is no indication of a contribution from the forward shock to the spectrum of SN~1993J. The absorption column density needed is in all cases just a little bit higher than the Galactic column density, which suggests that SN~1993J has significant contribution from the 
adiabatic reverse shock too, along with the radiative part.

We will now discuss the specific abundance zones one by one.
The best fit parameters for the solar composition, the CNO-enriched compositions from the model s15s7b2f \citep{Woosley94} and the SN~1987A-like composition are given in Table~\ref{fits:sn93j}. The spectra  were found to be best fit with one adiabatic component (denoted here by Ad. Comp) and one radiative component (Rad. Comp). For each component the table shows the best fit temperature, and the flux normalization constant (Norm) required to fit the spectrum.

\begin{table*}[h*]
\caption{Fitting parameters for different components for the SN 1993J XMM-Newton data. }
\label{fits:sn93j}
\begin{center}
\begin{tabular}{|ll|l|l|l|l|} 
\hline 
\multicolumn{2}{|c|}{} & \multicolumn{4}{|c|}{Models}\\
\cline{3-6}
\multicolumn{2}{|c|}{Parameters} & Solar & HHe-Zone & HeN-Zone & SN87A-like \\
\multicolumn{2}{|c|}{} &        & (s15s7)  & (s15s7)  &   \\
\hline
\multicolumn{2}{|c|}{$\chi^2$/{\it dof}} & 2.13 & 1.98 & 1.26 & 1.67\\
\hline
\multirow{2}{*}{Rad. Comp} & Temp (keV) &  $0.68\pm0.08$ & $0.71\pm0.08$ & 
$1.02\pm0.41$  & $0.75\pm0.11$\\[2pt]
                          & Norm (10$^{-3}$) &  $5.79\pm0.11$ & $6.53\pm0.83$ & 
$6.20\pm0.23$ & $5.10\pm0.73$\\
\hline 
\multirow{2}{*}{Ad. Comp} & Temp (keV) &  $1.33\pm0.24$ & $1.37\pm0.03$ & 
$2.07\pm0.16$  & $1.35\pm0.03$\\[2pt]
                          & Norm (10$^5$) &  $2.08\pm0.04$ & $1.81\pm0.05$ & 
$0.28\pm0.02$  & $1.58\pm0.03$\\
\hline
Unabsorbed flux $^*$ & Rad. Comp & $1.12\pm0.02$ & $1.24\pm 0.15$  & 
$1.64\pm 0.6$ & $1.14\pm 0.16$ \\[2pt]
$10^{-13}$ erg s$^{-1}$ & Ad. Comp & $1.72\pm 0.03$  & $1.59\pm 0.04$ & 
$1.45\pm 0.10$ & $1.73\pm 0.03$\\[2pt]
\cline{2-2}
  & Total flux & $2.84\pm 0.05$ &  $2.84\pm 0.19$& $3.09\pm 0.16$ & $2.87\pm 0.19$\\
\hline
\multicolumn{2}{|c|}{$n_{\mathrm{H}}$(cool shell) in $10^{20}$ cm$^{-2}$} &
$13.2\pm1.40$ & $11.8\pm0.96$ & $0.07\pm0.07$ & $6.5\pm1.1$ \\
\hline
Flux (Reverse Shock) $^{**}$ & Rad. Comp & $2.49\pm0.05$ & $2.46\pm 0.31$ & 
$2.73\pm 0.10$ & $2.18\pm 0.31$ \\[2pt]
$10^{-13}$ erg s$^{-1}$ & Ad. Comp & $2.60\pm0.05$  & $2.92\pm 0.08$ & 
$1.89\pm 0.14$ & $2.46\pm 0.05$\\[2pt]
\cline{2-2}
  & Total flux & $5.09\pm0.10$& $5.38\pm 0.39$ & 
$4.62\pm 0.24$& $4.66\pm 0.36$\\

\hline
\end{tabular}

$^*$ Corrected for the Galactic extinction column density, $n_{\mathrm{H}}$=
 $4.17\times10^{20}$ cm$^{-2}$\\
$^{**}$ The total flux from the reverse shock before absorption 
by the cool shell
\end{center}
\end{table*}

\subsection{A solar composition}
First we note that the relative strengths of the \ion{Mg}{xi} and \ion{Mg}{xii} lines, and of the \ion{Si}{xiii} and \ion{Si}{xiv} lines, indicate that there is a component with temperature in the range 1.0--2.0 keV. This is consistent with a density gradient $\eta \sim 10 - 12$. A lower temperature gives too strong  \ion{Mg}{xi}, and \ion{Si}{xiii} lines, while a higher temperature gives too strong \ion{Mg}{xii} and \ion{Si}{xiv} lines. However, temperatures above 1.0 keV strongly overestimate the emission at the Fe~L peak at 0.9 keV, while not having sufficient emission below 0.8 keV.  This may indicate that there is one component responsible for the line emission above 1.0 keV, and another, cooler component, responsible for the emission at lower energies.

For a solar composition the \ion{O}{viii}-emission at 0.6--0.7~keV is prominent. As this is absent in the observed spectrum, a strong absorption is needed to remove this emission. This, however, also removes the \ion{N}{vii}-emission at 0.5 keV, which is evident in the spectrum. 

\begin{figure*}[ht]
\begin{center}
\resizebox{0.6\hsize}{!}{\rotatebox{0}{\includegraphics{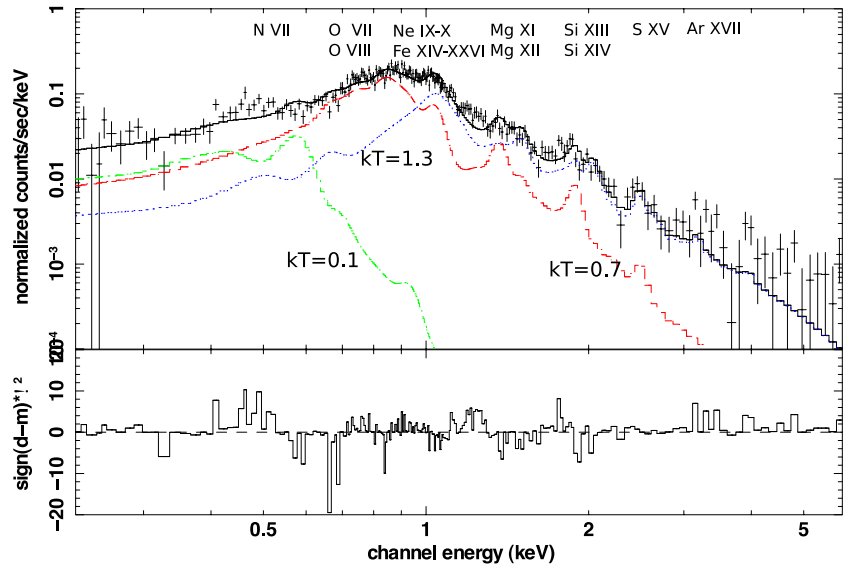}}}
\end{center}
\caption{Fit to the XMM spectrum of SN~1993J for an adiabatic shock with $kT_{\mathrm{rev}}=1.3$~keV  (dotted curve) and two radiative shock with $kT_{\mathrm{rev}}=0.7$~keV (upper dashed curve) and $kT_{\mathrm{rev}}=0.1$~keV (lower dashed curve) for a solar composition. The solid curve is the sum of the three components. Note the poor fit at the \ion{N}{vii} peak at $\sim 0.5$~keV and the \ion{O}{vii-viii} emission at $\sim 0.7$~keV.} 
\label{fig:93Jfit-solar-twoShocks}
\end{figure*}

We may improve this by adding a second component, which gives a best fit for shocks of $kT_{\mathrm{rev}}=1.0$ and $kT_{\mathrm{rev}}=2.0$ keV, shown in  Fig~\ref{fig:93Jfit-solar-twoShocks}. We note that while the XMM-observations are reasonably well reproduced in the range 0.8--1.0 keV, the Chandra observations show a dip at 0.9 keV, which means that this model overestimates the emission there. Both spectra are poorly reproduced below 0.7 keV, most likely because of the strong absorption needed to remove the \ion{O}{viii}-emission. 

We therefore conclude that a composition with less O and more N than the solar composition is needed. This is in agreement with the results from the optical emission, which indicate the presence of CNO-processed material \citep{Fransson05}. In the following we therefore investigate the spectra produced by compositions taken from the explosion models described above.

\subsection{A CNO enriched hydrogen envelope}
The composition from the outer envelope of the 4H47 model, where the original envelope composition has been modified by CNO-processing in the progenitor (Table~\ref{tab:abund}), gives weaker O-lines than the solar model, while the N-lines are stronger. However, also this composition gives too weak  \ion{N}{vii} emission at $\sim 0.5$ keV. These models also show a deficiency at 0.6--0.7~keV compared to the observed spectrum, while the Fe~L/\ion{Ne}{x} peak at 0.9--1.0 keV is too strong. This indicates that this composition has too much Ne and too little O. At energies above $\sim 1.0$ keV the continuum is too weak, and there is hardly any line emission. 

\begin{figure*}[ht]
\begin{center}
\resizebox{0.6\hsize}{!}{\rotatebox{0}{\includegraphics{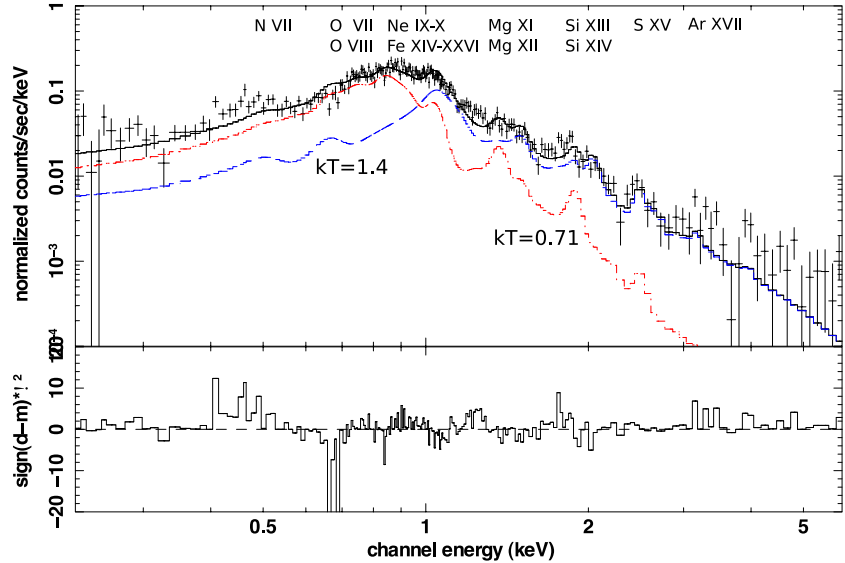}}}
\end{center}
\caption{Fit to the spectrum of SN~1993J with an adiabatic shock at $kT_{\mathrm{rev}}=1.4$~keV  (long dashed) and a radiative shock at $kT_{\mathrm{rev}}=0.71$~keV (short dashed) for a composition based on the outer, hydrogen-rich envelope of the s15s7b2f model. Also here the fit to the \ion{N}{vii} and \ion{O}{vii-viii} emission at 0.5-0.7 keV is poor.} 
\label{fig:93Jfit-HHezone}
\end{figure*}

The outer envelope of the s15s7b2f model has more O and less Ne than the corresponding zone in the 4H47 model, giving stronger \ion{O}{vii-viii} lines and a less pronounced \ion{Ne}{x} peak. Also, the \ion{Fe}{xvii} lines are stronger, as are the \ion{Mg}{xi-xii} and \ion{Si}{xiii-xiv} lines at 1.2 keV and 1.8 keV. However, although this model does a better job at reproducing the Fe~L emission and the line emission above 1.2 keV, also in this model the O-emission at 0.7--0.8 keV is underestimated, and the Ne-peak is too strong. In this case two shocks improve the fit, but cannot reproduce the missing O-emission. Fig.~\ref{fig:93Jfit-HHezone} shows a two-shock model based on the model s1517b2f. In this case the temperatures of the two shocks are 0.7 and 1.4 keV.

As a realistic CNO-enriched composition we have also tested the SN~1987A-like composition discussed in Sect.~\ref{sect:results}. The fit obtained with this composition is better than with the other two CNO-enriched compositions, although the \ion{N}{vii} peak at 0.5 keV also in this case is underproduced, while the \ion{Fe}{xvii} emission at 0.8 keV is too strong.

\begin{figure*}[ht]
\begin{center}
\resizebox{0.6\hsize}{!}{\rotatebox{0}{\includegraphics{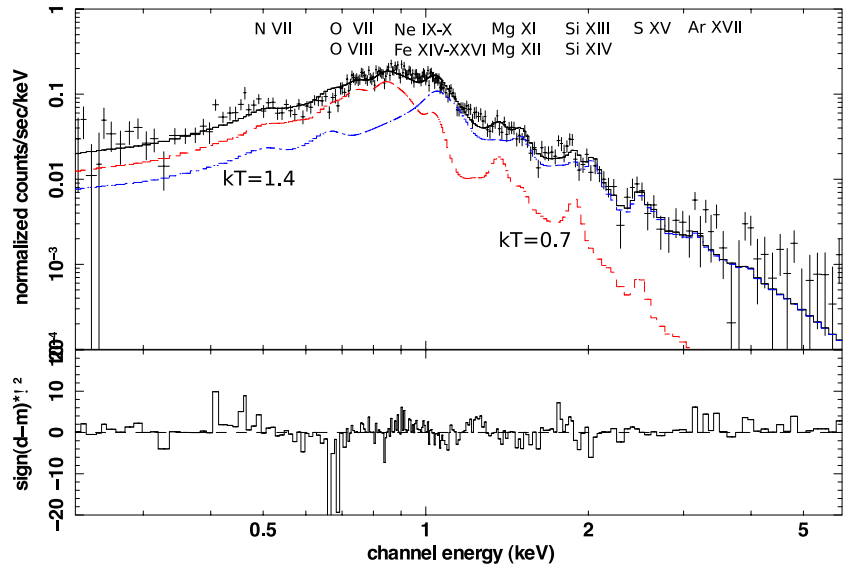}}}
\end{center}
\caption{Fit to the XMM spectrum of SN~1993J with a composition resembling that of the circumstellar ring of SN~1987A, but with enhanced metal abundances. The fit consists of an adiabatic shock at $kT_{\mathrm{rev}}=1.4$~keV (dotted) and a radiative shock at $kT_{\mathrm{rev}}$=0.7~keV (dashed). The fit to the \ion{N}{vii} peak at $\sim 0.5$~keV has improved somewhat compared to Figs~\ref{fig:93Jfit-solar-twoShocks}-\ref{fig:93Jfit-HHezone}, but the \ion{O}{vii-viii} emission at 0.7 keV is still poorly fit.} 
\label{fig:93Jfit-87Alike}
\end{figure*}

Adding a second shock improves the fit further by allowing strong line emission at high energies without overestimating the Fe emission at 0.9 keV. An example, with two radiative shocks with $T_{1}=1.4$~keV and $T_{2}=0.7$~keV, is shown in Fig.~\ref{fig:93Jfit-87Alike}.

In summary, this composition gives an acceptable fit to the observations.

\subsection{The Helium/Nitrogen zone}
This zone, dominated by CNO burning products, with He dominant and N and Ne the most abundant metalsreproduces the emission around 0.5 keV well. The Fe~L and \ion{Ne}{x} peak at 0.9 keV is, on the other hand, overestimated, and there is an obvious lack of emission at 0.6-0.7 keV, due to a too low abundance of O. The s15s7b2f model produces an Fe~L/\ion{Ne}{x} peak at slightly lower energy than the 4H47 model, due to the higher \ion{Ne}{x}-emission in the 4H47 model. Several emission lines are evident above 1 keV, stronger in the s15s7b2f model. 

\begin{figure*}[ht]
\begin{center}
\resizebox{0.6\hsize}{!}{\rotatebox{0}{\includegraphics{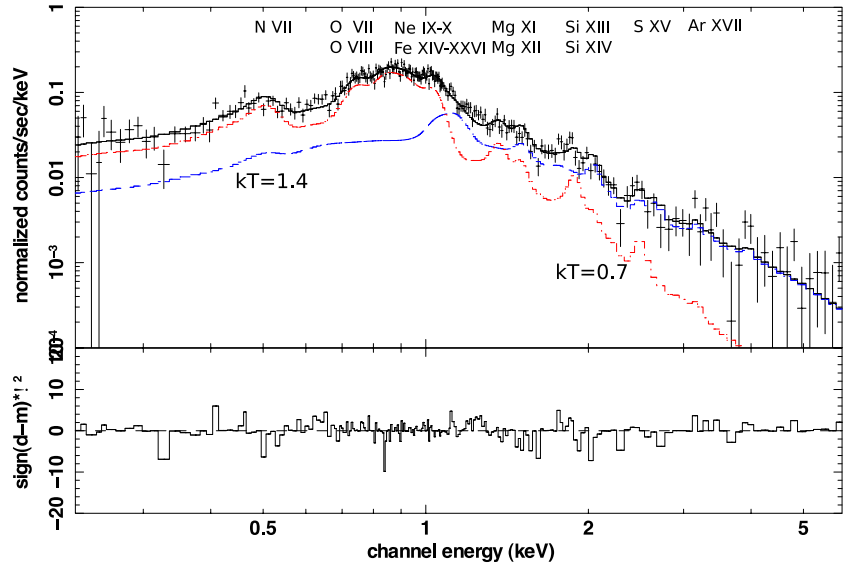}}}
\end{center}
\caption{Fit to the XMM spectrum of SN~1993J with an adiabatic shock at $kT_{\mathrm{rev}}=2.1$~keV  (long dashed) and a radiative shock at $kT_{\mathrm{rev}}=1.0$~keV (short dashed), for a He/N-dominated composition from the s15s7b2f model. Note the strong improvement in the fit to the \ion{N}{vii} peak at 0.5~keV compared to Figs.~\ref{fig:93Jfit-solar-twoShocks}--\ref{fig:93Jfit-87Alike}.} 
\label{fig:93Jfit-HeNzone}
\end{figure*}

As with the previous models, adding a second shock improves the fit, and the best fit is found for an adiabatic shock at $kT_{\mathrm{rev}}=2.1$~keV and a radiative component at $kT_{\mathrm{rev}}=1.0$~keV (Fig.~\ref{fig:93Jfit-HeNzone}). This gives the best fit to the spectrum of all the compositions. 

The highest temperature component is consistent with an ejecta density gradient of $\eta \sim 12$ if $\tilde{A_{*}}=4$ and $V_{4}=1$ (Table~\ref{tab:limTemp}).

\subsection{The Carbon/Oxygen zone}
\begin{figure*}[ht]
\begin{center}
\resizebox{0.6\hsize}{!}{\rotatebox{0}{\includegraphics{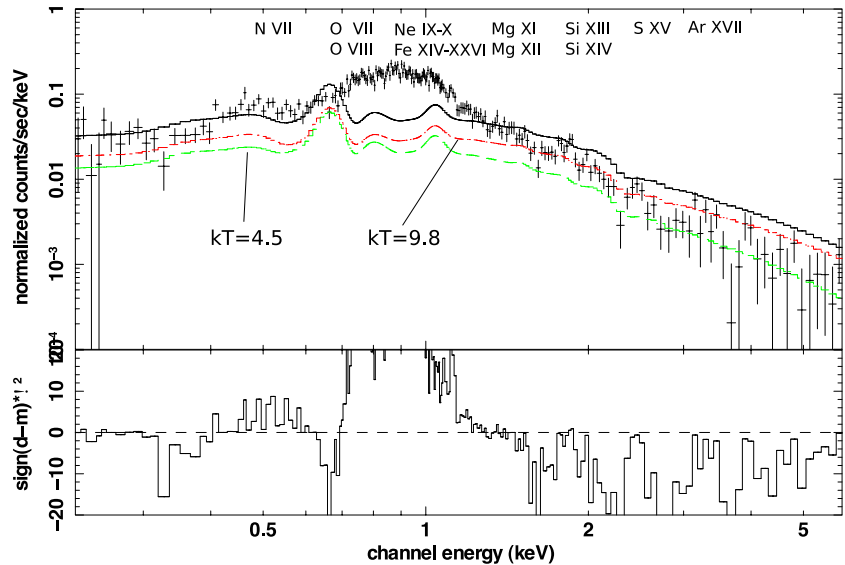}}}
\end{center}
\caption{Fit to the XMM spectrum of SN~1993J with two radiative shocks with $kT_{\mathrm{rev}}=4.5$~keV (lower dashed curve) and $kT_{\mathrm{rev}}=9.8$ (upper dashed curve), for a C/O-dominated composition from the 4H47 model. Note the poor fit at 0.7-1.2 keV, ruling out this model.} 
\label{fig:93Jfit-CO}
\end{figure*}

A C/O-dominated composition gives spectra that do not in any way resemble the observed spectrum. For illustration we show one in Fig.~\ref{fig:93Jfit-CO}. Both the N peak at 0.5 keV Fe peak at 0.9 keV are completely missing, while the oxygen emission at 0.6-0.7 keV and the Mg-S line emission above 1 keV are very strong and dominate the spectrum. Changing the temperature, or adding more shocks, modifies the strength of the peaks, but does not result in an overall improvement.

\section{Discussion}
\label{sect:discussion}
In this paper we have investigated the X-ray spectra of SN~1993J. By applying a hydrodynamical model for a cooling shock to the X-ray observations we have discussed the fits obtained by various compositions taken from realistic ejecta models. 

We find that in the spectra of SN~1993J roughly half the contribution comes 
from  an adiabatic shock propagating into a CNO-enriched ejecta, with the rest of
the contributions from slower, radiative shocks. These could be caused by instabilities at the reverse shock front or by a clumpy CSM modifying the structure of the interaction region. Comparing ejecta models, we find that the model s15s7b2f \citep{Woosley94} gives a somewhat better fit to the line features than the model 4H47 \citep{NomHash88, Shigeyama94}, but the difference is hardly significant. The best fits are obtained by a composition resembling that of the circumstellar ring of SN~1987A. There is no need for a contribution from the circumstellar shock, as was also expected from the low luminosity estimated in Sect. \ref{sect:csEmission}.

The observations discussed in the present work are close to the limit of what can be used for spectroscopic studies. If the capabilities of Chandra and XMM had been present at the early phases of this supernova, a much more detailed analysis could have been performed of the X-ray spectra. The next X-ray bright supernova within $\sim 10$~Mpc will allow a much more detailed spectroscopic studies of the early spectra. With future missions, like ESA's, NASA's and JAXA's proposed International X-ray Observatory (IXO), these capabilities will be vastly improved. 
This will allow high S/N spectra to be obtained of SN at much longer distances, and will also allow detailed spectroscopic studies to be performed also at late times. This will in turn lead to a much more reliable determination of the abundances and temperature in the interaction region. We hope that the analysis in this paper will give some guidance to future X-ray observations of SNe.

\section{Acknowledgements}
We are grateful to Stefan Larsson and Linnea Hjalmarsdotter for assistance with XSPEC and to Ken Nomoto and Stan Woosley for detailed abundance models of SN~1993J. This work has in part been supported by the Swedish National Space Board and the Swedish Research Council. P.C. is a Jansky fellow at the National Radio Astronomy Observatory, which is a facility of the National Science Foundation operated under cooperative agreement by Associated Universities, Inc.

\bibliography{0884}
\bibliographystyle{aa}

\end{document}